\documentclass[preprint]{aastex}

\begin{document}

\slugcomment{Submitted to Publ. Astro. Soc. Pacific,  January 2013}

\title{VOStat: A Statistical Web Service for Astronomers}
\shorttitle{VOStat Web Service}

\author{
Arnab Chakraborty\altaffilmark{*} \altaffilmark{1,2}, 
Eric D. Feigelson\altaffilmark{1,3,4}, 
G. Jogesh Babu\altaffilmark{}\altaffilmark{1,4,3}
}

\altaffiltext{*}{Send requests to A. Chakraborty at arnabc@isical.ac.in}
\altaffiltext{1}{ Center for Astrostatistics, 417 Thomas Building, Pennsylvania State University, University Park PA 16802}
\altaffiltext{2}{Applied Statistics Unit, Indian Statistical Institute, 203 Barrackpore Trunk Road, Kolkata 700108, India}
\altaffiltext{3}{Department of Astronomy \& Astrophysics, 525 Davey Laboratory, Pennsylvania State University, University Park, PA 16802} 
\altaffiltext{4}{Department of Statistics, 325 Thomas Building, Pennsylvania State University, University Park PA 16802}

\begin{abstract}
VOStat is a Web service providing interactive statistical analysis of astronomical tabular datasets.  It is integrated into the suite of analysis and visualization tools associated with the international Virtual Observatory (VO) through the SAMP communication system.  A user supplies VOStat with a dataset extracted from the VO, or otherwise acquired, and chooses among $\sim 60$ statistical functions.  These include data transformations, plots and summaries, density estimation, one- and two-sample hypothesis tests, global and local regressions, multivariate analysis and clustering, spatial analysis, directional statistics, survival analysis (for censored data like upper limits), and time series analysis. The statistical operations are performed using the public domain {\bf R} statistical software environment, including a small fraction of its $>4000$ {\bf CRAN} add-on packages.  The purpose of VOStat is to facilitate a wider range of statistical analyses than are commonly used in astronomy, and to promote use of more advanced methodology in {\bf R} and {\bf CRAN}.  
\end{abstract}


\section{Introduction}
Astronomy has always accounted for the production of a significant amount of
scientific data. With recent developments in instrumentation,
this amount surpassed most fields of
science. Most of the astronomers, however, are still using
traditional techniques to analyze the available data. While this
approach 
has definitely produced many good results, it is expected to be
even more productive if it is coupled with statistical
techniques: the time-tested science of data analysis. Both
astronomy and statistics are well developed branches of science,
each with its own set of journals, conferences, degrees and
jargons that require time and patience to master. So desirable though 
it is, simultaneous proficiency in both the fields  not commonly achieved.

VOStat\footnote{VOStat is available at \url{http://vostat.org}.} is a web service that is designed to help in this endeavor. 
Oriented towards tabular data, rather than images or spectra, it is an effort to bring statistics to the astronomer's
desktop. There seems to be two ways to introduce statistics to an
astronomer. The first is to ask her to read
and digest a selection of standard books on
statistics (most of which are written with social scientists in
mind), and then expect her to apply the ideas to problems in her
own research. Statistics texts designed for the physical scientist are
now emerging to assist in this process \citep{James06, Wall12, Feigelson12}.  
The second approach is to offer her a convenient Web service with a menu of
statistical methods to choose from. Data can be uploaded from her local
computer or acquired from the Virtual Observatory, a worldwide federation 
of public astronomical databases and associated analysis services.  
She can pick methods and immediately
apply them to her data without having to download software or read
more than a short non-technical description. If she
likes the result, she may then delve deeper into the methodology 
and acquire the underlying software to investigate the initial findings
more carefully on her machine. 

VOStat takes the latter approach. It uses {\bf R}, the largest free, public
domain statistical software to perform its computations.  The
output of VOStat is not merely the output of {\bf R}, but also the
R scripts (generated automatically by VOStat based on the user's
specification).  Indeed, the main goal is not to encourage
astronomers to outsource their statistical requirements to
VOStat, but to teach them how to be able to carry out independent
statistical analysis of their data themselves.  VOStat is available to 
anyone on the World Wide Web, but is particularly oriented towards
astronomers drawing upon Virtual Observatory datasets and tools.

\section{The {\bf R} Statistical Software Environment}

For most of the computer age, the primary software implementing statistical analyses were commercial packages.  The Statistical Analysis System\footnote{\url{http://www.sas.com}}, providing a vast array of built-in procedures in a Fortran-like programming language, is the most comprehensive, serving industries and governments.  During the 1980s, John Cambers and colleagues at AT\&T developed {\bf S} \citep{Chambers98}, a statistical programming language written in C, that  evolved into  the commercial package {\bf S-Plus}\footnote{\url{http://spotfire.tibco.com/products/s-plus/statistical-analysis-software.aspx}}.  In the 1990s, New Zealand statisticians Ross Ihaka and Robert Gentleman rewrote {\bf S} in the public domain, calling it {\bf R}, and released it under GNU General Public License software product under the auspices of a non-profit {\bf R} Foundation for Statistical Computing led by a growing {\bf R} Development Core Team \citep{Ihaka96}.  

In the spirit of open source software, the {\bf R} team set up the Comprehensive {\bf R} Archive Network ({\bf  CRAN} ), a structure for external experts in statistical computing to contribute specialized packages.  About 20 of the early packages were incorporated into the base {\bf R} package, which is now a reasonably stable product.  But {\bf  CRAN}  contributions continued to enter the system, growing exponentially since 2001.  Today there are $>4000$ {\bf  CRAN}  packages with a new package entering every day.  They cover every field of applied statistics with particularly strong contributions from the biology (particularly genomics) and econometrics communities.  The {\bf R/ CRAN}  system now has $>60,000$ statistical functionalities, some simple and others very complex.  {\bf  CRAN}  packages are easily retrieved on-the-fly during an {\bf R} session.  

As there is no global index to {\bf  CRAN} , it can be tricky to find the appropriate functionality for the task at hand.  The {\bf  CRAN}  Task Views\footnote{\url{http://cran.r-project.org/web/views}} provide brief but useful summaries of packages in $\sim 30$ broad topics.  These include Bayesian inference, chemometrics and computational physics, cluster analysis and finite mixture models, graphic displays and visualization, high performance computing, machine learning, medical imaging, multivariate analysis, optimization and mathematical programming, robust statistical methods, spatial data, survival analysis (treating upper limits), and time series analysis.  Publications using {\bf R/ CRAN}  should cite \citet{R12} and the reference given by the function {\it citation\{package\}}.  

The capabilities of VOStat are similar in scope to those provided by other {\bf R} graphical user interfaces (GUIs) like {\it Deducer} and is less extensive than some like {\it  R Commander}\footnote{See \url{http://www.sciviews.org/\_rgui} for details on $\sim 25$ GUIs developed for {\bf R}.}.  The purpose for creating a new {\bf R} GUI for astronomy is interoperability with other Virtual Observatory capabilities through its Simple Application Messinging Protocol (SAMP), so the astronomer experiences a flexible and capable environment for  interactive scientific  analysis of tabular data\footnote{Virtual Observatory science software tools and services with SAMP connectivity are available at \url{http://www.usvao.org/science-tools-services/}, \url{http://www.euro-vo.org/pub/fc/software.html}, and \url{http://wiki.ivoa.net/twiki/bin/view/IVOA/SampSoftware}.}.

\section{Overview of VOStat}
The international Virtual Observatory (VO) system federates the data resources of various astronomical facilities around the world.  It provldes tools for the extraction, visualization, and comparison of data extracted from VO databases, as well as some specialized astronomical analyses such as construction and fitting spectral energy distributions and lightcurves for variables objects.   This system provides easy web-access to a vast amount of astronomical data, and allows astronomers to communicate findings easily for multi-observatory studies. Its impressive scope notwithstanding, the VO system lacks comprehensive statistical tools suitable for analyzing the data extracted by a user. 

The VOStat project addresses this deficiency with two goals: it provides a statistical toolbox well integrated into the VO system to enable extracted astronomical data to be analyzed efficiently and effectively; and it spreads awareness among practicing astronomers about  the need of complementing domain knowledge with statistically valid data analysis techniques.  VOStat seeks to achieve these aims with a web-based astrostatistical computing portal that links VO-extracted datasets to the statistical analysis capabilities of {\bf R}, the largest free, open-source statistical computing environment.   Automated connectivity between these software environments is provided by the VO's Simple Application Management Proocol (SAMP), a VO astronomer can (for example) interactively acquire tabular datasets from the VO Data Discovery Tool, view and manipulate the tables using TOPCAT or VOTable,  and perform statistical analyses with graphics using VOStat.    While the analyses provided by VOStat are not highly complex due to the limitations of the Web services interface, {\bf R} code is provided giving a springboard for the astronomer to develop and implement more advanced statistical methodology geared towards the particular astronomical research problem at hand. 

VOStat is accessed by the public at {\tt www.vostat.org}.   Astronomers can upload their data into the VOStat server via a simple Graphical User Interface that operates the chosen {\bf R} functions on the user-provided dataset.  The data files can be on the user's home computer, on a Web site, or obtained from VO-compliant tools via the VO's  SAMP.   VOStat thus can interact with human users directly (e.g., accepting data, displaying plots on screen), with VO-compliant tools via a SAMP hub, a specialized server running in the local machine of the user that uses SAMP as its communication protocol. 

Once a dataset is extracted from the VO or otherwise provided, the user chooses from a list of several dozen statistical techniques, the selected techniques are applied to the data, and the results are returned to the user. Apart from traditional output like on-screen plots and tables, VOStat also provides the following to understand and extend the statistical analysis of their dataset on their home computer: 
\begin{enumerate}
\item   annotated {\bf R} code for each step of the statistical analysis;
\item   workspace file giving internal {\bf R} files of statistical analysis results;
\item   help files from {\bf R} documentation and from Wikipedia describing the statistical functions;
\item   guidance for the statistically uninitiated astronomer to select the right analysis.
\end{enumerate}

VOStat is oriented towards analysis of tabular data.  Commonly, the rows of the table represent a collection of astronomical objects and the columns represent measured or inferred properties of the objects.  But  astronomical lightcurves or spectra can be analyzed, where the first column gives observed times or wavelengths and the other columns give observed values.  {\bf R} has extensive capabilities for analysis of images (see the {\bf  CRAN} Task View on medical image analysis, http://cran.r-project.org/web/views/MedicalImaging.html), but these are not incorporated into VOStat.  Therefore, we do not encourage analysis of astronomical images using VOStat.  

A common concern among astronomers while choosing a software 
is the upper bound of data size that can be handled. VOStat
imposes no further restriction on this beyond the limitation of R
itself. Thanks to the way R is built from diverse packages
created by unrelated groups of developers, it is difficult to
quantify the data size limitations of R in general. Certain
methods allow quite large data sets, while others do not. The
primary design goal behind VOStat is to motivate astronomers to
use R. So users are encouraged to play with the available methods
using moderately sized subsets of their data, and then download and tweak the generated
script to work with the entire data sets. VOStat has facilities
built into it to choose a subset of data.

Two earlier versions of VOStat were developed, the first produced by a U.S. group involving Penn State and Caltech, and another undertaken by VO-India and  Penn State with contributions from Caltech and Calcutta University Statistics Department. The latter is provided for use by VO-India\footnote{\url{http://vo.iucaa.ernet.in/$\sim$voi/VOStat.html}}.  These versions of VOStat did not have SAMP communication with other VO tools and provided a narrower suite of statistical methods. Other tools for interfacing {\bf R} to the Web for general statistical analyses without association with the Virtual Observatory include FastRWeb, R2HTML, Rcgi, Rweb and Rwui\footnote{\url{http://www.rforge.net/FastRWeb}, \url{http://cran.r-project.org/web/packages/R2HTML}, \url{http://www.ms.uky.edu/$\sim$statweb}, \url{http://www.math.montana.edu/Rweb}, \url{http://sysbio.mrc-bsu.cam.ac.uk/Rwui}}. 

\section{Internal structure of VOStat} 

VOStat uses Java as the interface between {\bf R} and the user.  The Apache Tomcat web server deploys the Java servlets that comprise the web interface.  An {\bf R} session is then spawned on the server computer to implement the chosen statistical procedures.  Figure~\ref{fig:vosflow} shows the workings of VOStat, and the three principal modules are described here.

\begin{figure}
\includegraphics[width=0.9\textwidth]{f1.eps}
\caption{Flowchart for VOStat operation.
\label{fig:vosflow}}
\end{figure}

\subsection{Data loader module}
VOStat allows data to be fed into it from different sources:
\begin{enumerate}
\item 	Uploading a file from the user's local machine.
\item 	Uploading a file from a URL on the Internet.
\item	Getting data from VO-compatible tools via a SAMP hub. These tools include Aladin Sky Atlas, Datascope, Octet, Open SkyQuery, Topcat, VIM, and VisIVO. 
\end{enumerate}

We rely on Java applet technology for the SAMP communication, and on multipart file uploading for accessing a file from the user's local disk.  The data file may be in different formats commonly in use among astronomers: VOTable, FITS, or ASCII tab/space/comma separated values.   VOStat can guess the file format from the file name extension, or the user may specify the file format from a drop-down menu.  The file extensions recognized by VOStat are .txt, .dat, .csv, .fits, .vot and .xml.   These formats are read using {\bf R} utilities. 

For datasets available on the Web, the first page of VOStat has a text field where the user can specify the URL of the file to be uploaded. A similar interface allows files to be uploaded from the user's hard disk.  Obtaining data via SAMP from VO-compliant tools is slightly more involved, because the SAMP hub must run in the user's local machine while VOStat resides in a remote server. For this we embed a trusted Java applet in the webpage served by the VOStat server. This applet contains two clients in it, a SAMP client and an HTTP client. The SAMP client looks for and connects to any SAMP hub found running in the local machine. Once a file broadcast message (MType: table.load.votable) is received by the SAMP client, the HTTP client establishes connection with the remote VOStat server. The file is read by the applet and sent over the internet to a servlet running in the VOStat server. This indirect process is required because, during 2011-12 when VOStat was developed, a SAMP hub can run in the local machine while VOStat tables are broadcast via local URLs only. 

\subsection{Analysis Module}
This is written entirely in R.  For the more commonly used statistical functions, the code uses functions in base R.  For  more specialized functions, relevant  {\bf  CRAN} packages are first installed on the server machine and the {\bf  CRAN} functions are run on the dataset. In a few cases, we implement our own analysis techniques designed for astronomers. The VOStat architecture is created with the design goal that new statistical techniques may be added easily without disrupting the existing workflow.  The analysis tools currently in VOStat are briefly described in \S\ref{stat_fns.sec}.

\subsection{Output Module}

Three types of output from VOStat are provided.  First, various plots, estimated parameter values, confidence intervals, correlation coefficients, diagnostic messages, and other ASCII results are for direct use by the user.  Numerical output is usually short and appears on the screen. Plots can be downloaded in PDF or Encapsulated PostScript formats.  Second, tables meant for consumption by other SAMP-aware VO tools are broadcasted through the local SAMP client.  Third, the {\bf R} binary workspace (usually called {\it .Rdata}) is available for download so that an analysis can be continued seamlessly at a later session. This can assist users who want to continue more varied and advanced using {\bf R} installed in their own machines.

To further the goal of educating users about statistical tools, VOStat also gives background information about the performed analysis.  First, annotated {\bf R} codes can be `cut and paste' to reproduce the analysis on a home computer.  The VOStat computation is thus not an opaque `black box' but can be reproduced and extended by the scientist.  Second, links are provided to the {\bf R} documentation for each important {\bf R} function used.  These `help files' have a standard form defining the function, presenting the inputs and outputs, algorithms, and generic examples of their use.  Third, links are given to Wikipedia entries to give a more detailed and mathematical presentation of the statistical method.  References to the methodological literature are given in both the {\bf R} help files and Wikipedia entries.  Nearly all of the statistical functions in VOStat are also described, with applications to astronomical datasets, in the textbook {\it Modern Statistical Methods in Astronomy with {\bf R} Applications} \citep{Feigelson12}.  This text provides 19 astronomical datasets that can be used as exercises for VOStat application and more elaborate R code illustrating statistical analyses of these datasets\footnote{\url{http://astrostatistics.psu.edu/MSMA/datasets} and \url{http://astrostatistics.psu.edu/MSMA/MSMA\_R\_scripts.html}}.

For analyses that produce new data worth feeding into other VO-tools (for example, for advanced visualization) can be  routed via a SAMP hub running in the user's machine. Like other web-based VO-tools, VOStat uses trusted applets to connect to a SAMP hub running in the local machine. However, unlike some other tools like Aladin Sky Atlas, VOStat does not start a local hub itself. It only connects to a hub that is already running. This is to avoid disrupting coordination of the local VO-tools in case of a browser failure.

\section{VOStat statistical functions}
\label{stat_fns.sec}

VOStat (Rev. 2, December 2012) currently provides over 50 statistical functionalities organized into 11 topics.  We review these briefly here.  {\bf R} and {\bf  CRAN}  functions are given in brackets and italics, respectively.  Modules that produce output that can be automatically transferred to another VO-compatible client via SAMP are labeled {\sc `SAMP'}.  Figures of selected graphical outputs were generated in a separate session of {\bf R} from the VOStat {\bf R} codes in order to cosmetically improve them for publication (e.g., axis labels).  Intermediate-level descriptions of these statistical functions with applications to astronomical data are given by \citet{Feigelson12}.  

\subsection{Transforming and subsetting the data} \label{funct_transf}

{\bf Data summary} ~~ This gives minimum, quartiles (25\%, 50\%, 75\%), and maximum values for each variable. [{\it summary}] 

{\bf Select cases}~~ Rows can be selected using operations $<$, $\le$, $>$, and $\ge$ applied to specified columns. 

{\bf Simple transform}~~  Variables can be transformed with
$\log_{10} x$, $\sqrt{x}$, $\sqrt[3]{x}$, $x^2$,  $x^3$, $1/x$
and standardization (centering and scaling).

\subsection{Plots and summaries}  \label{funct_plots}

{\bf Boxplot} ~~ Boxplots show the minimum, maximum, median and quartiles of a univariate distribution with tails and outliers \citep{Tukey77}. [{\it boxplot}] 

{\bf Dotplot} ~~  Dotplots show univariate values of individual
objects with labels. This is appropriate (in terms of visual
clarity) only for small datasets.  [{\it dotchart}] 

{\bf Histogram} ~~ Histograms show the univariate distribution grouped into bins.  The user can specify the number of bins or the bin boundaries. If none are given, then the number of bins uses the Freedman-Diaconis algorithm based on the inter-quartile range \citep{Freedman81}.   [{\it hist}] 

\begin{figure}
\centering
\includegraphics[width=0.45\textwidth]{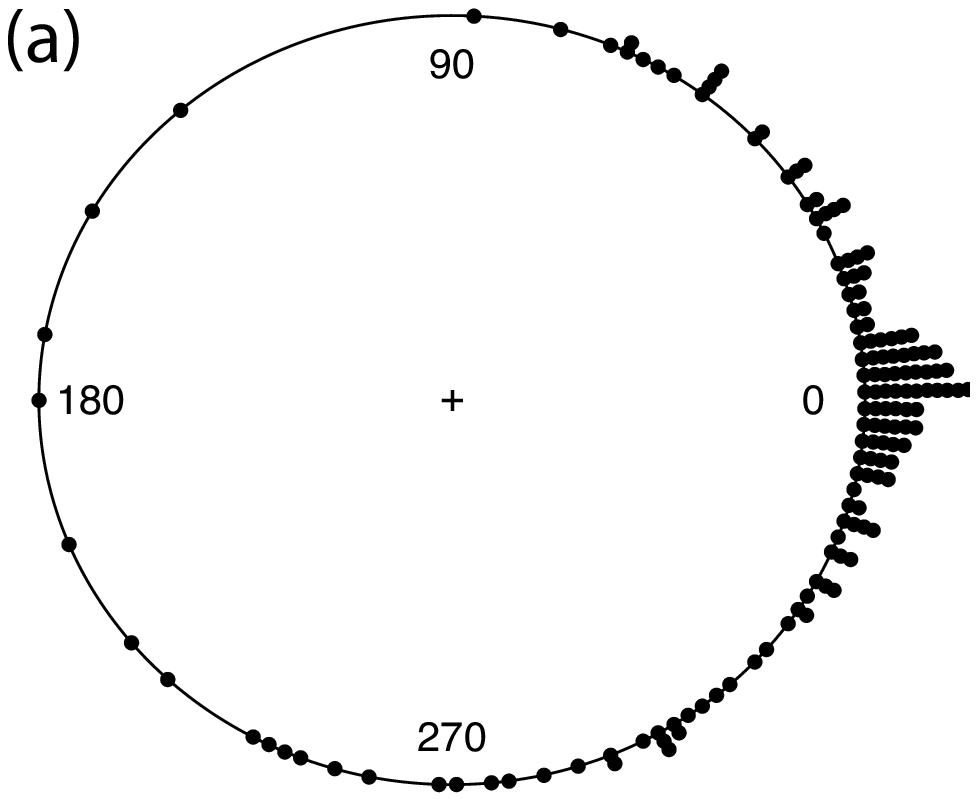}
\includegraphics[width=0.45\textwidth]{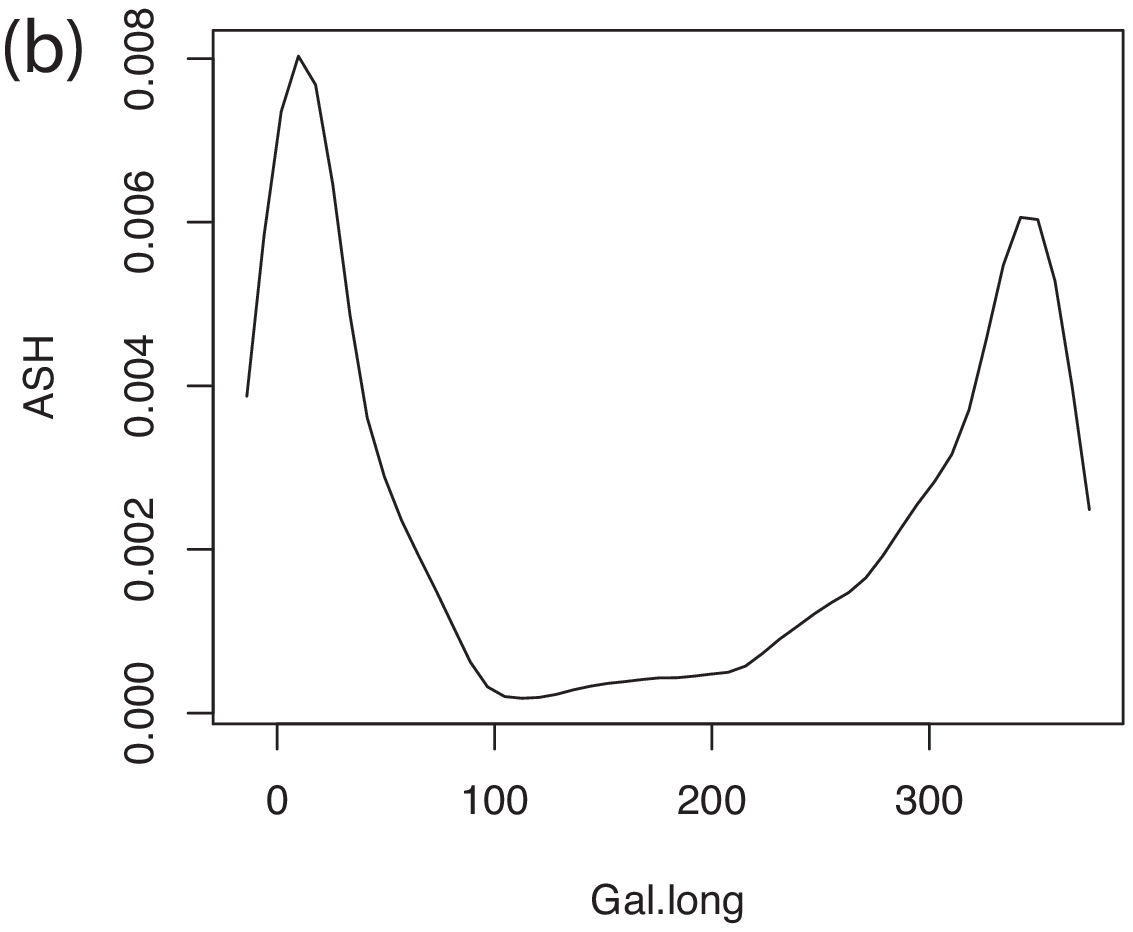} \\
\includegraphics[width=0.6\textwidth]{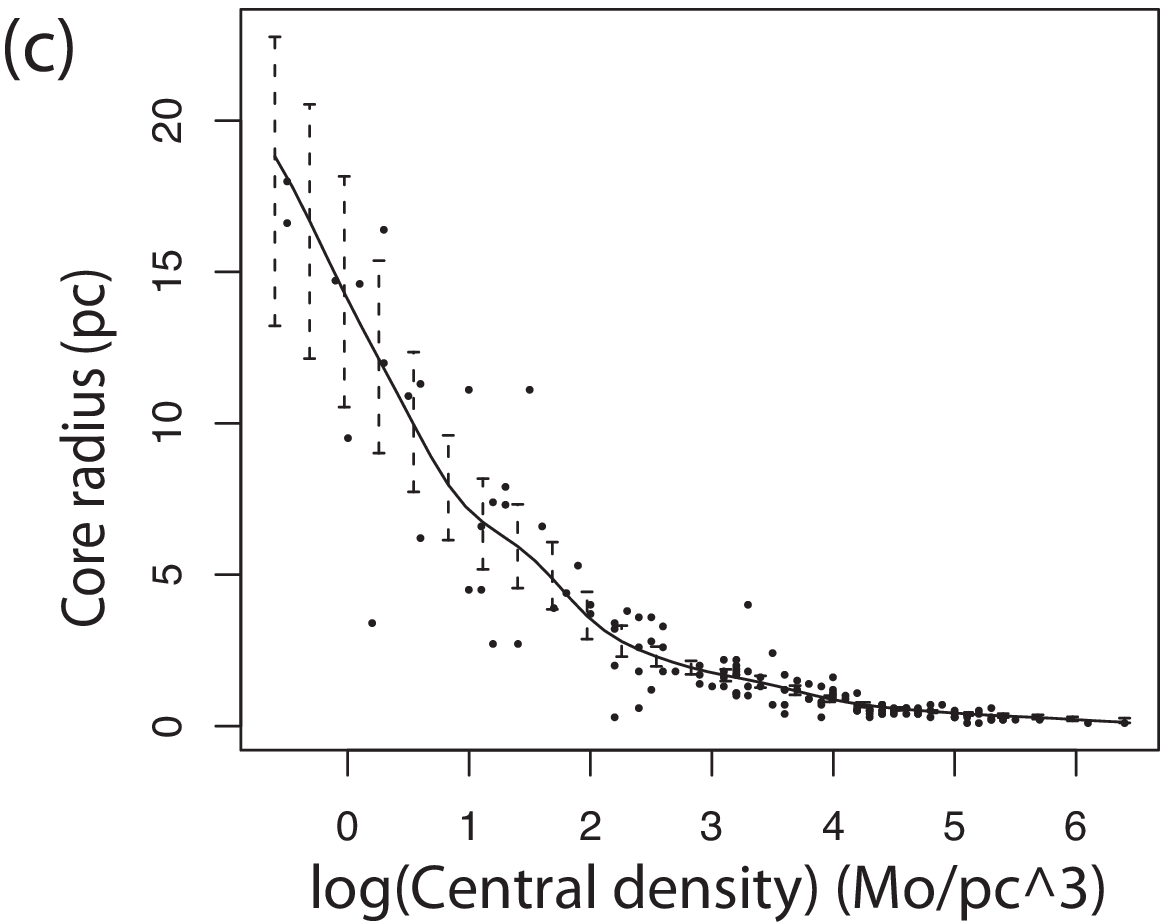} 
\caption{Selected VOStat plots using a multivariate dataset of properties of Galactic globular clusters \citep{Webbink85} available at http://astrostatistics.psu.edu/MSMA/datasets/GlobClus\_prop.dat. Plot formats have been adjusted in {\bf R} for publication.   (a) Directional plot of Galactic longitudes. (b) Averaged shifted histogram of cluster Galactic longitudes.  (c) Nadaraya-Watson local regression of globular cluster core radius (in pc) on the log of the central stellar density with 95\% bootstrap confidence intervals. (d) Pairs plot showing bivariate relationships between four properties of globular clusters (visual magnitude, core radius, concentration and log central star density).  (e) Pair (2-point) correlation function of globular cluster Galactic longitudes and latitudes (black), with an edge correction (red) and a theoretical line assuming random locations (green). (f) Voronoi tessellation of globular cluster visual magnitude and log central star density. (g) Adaptive smoothed image based on Voronoi tessellation.  
\label{multivar_plots.fig}}
\end{figure}

{\bf Averaged shifted histogram} ~~ Averaged shifted histograms are univariate density estimators (smoothers) that repeatedly bins the data with shifted origins and convolved with a quartic (biweight) kernel \citep{Scott92}.  The user specifies the number of bins used.  This is implemented in the {\bf  CRAN} {\it ash} package \citep{Scott12}.  An example is shown in Figure~\ref{multivar_plots.fig}b.  [{\it ash1}] 

{\bf Scatterplot} ~~ This is the common plot of individual data points in two variables specified by the user. Numerous options in formatting this plot (and other VOStat graphical outputs) are available to the user who adjusts the plot parameters in {\bf R} on their home computer.  [{\it plot}] 

{\bf Pairs plot} ~~  A pairs plot is a matrix of scatterplots for a multivariate dataset, sometimes called a SPLOM (for `Scatter PLOt Matrix').    It is often valuable for examining patterns and locating problems in multivariate data.  In practice, a limited number of variables (say $<10$) can be shown in a single plot. An example is shown in Figure~\ref{multivar_plots.fig}d.  [{\it pairs}] 

{\bf 3D scatterplot} ~~ This is a plot of three variables specified by the user implemented in {\bf  CRAN} package {\it scatterplot3d} \citep{Ligges03}. [{\it scatterplot3d}] 

{\bf Summary statistics} ~~ This gives mean, median, variance, median absolute deviation for each variable specified by the user.  It also shows a matrix of Kendall's $\tau$ nonparametric coefficient of bivariate correlation with associated probability values obtained with {\bf  CRAN} package {\it SuppDists} \citep{Wheeler09}.  [{\it mean, median, var, mad, cor, pKendall}] 

\begin{figure}
\centering
\includegraphics[width=0.50\textwidth]{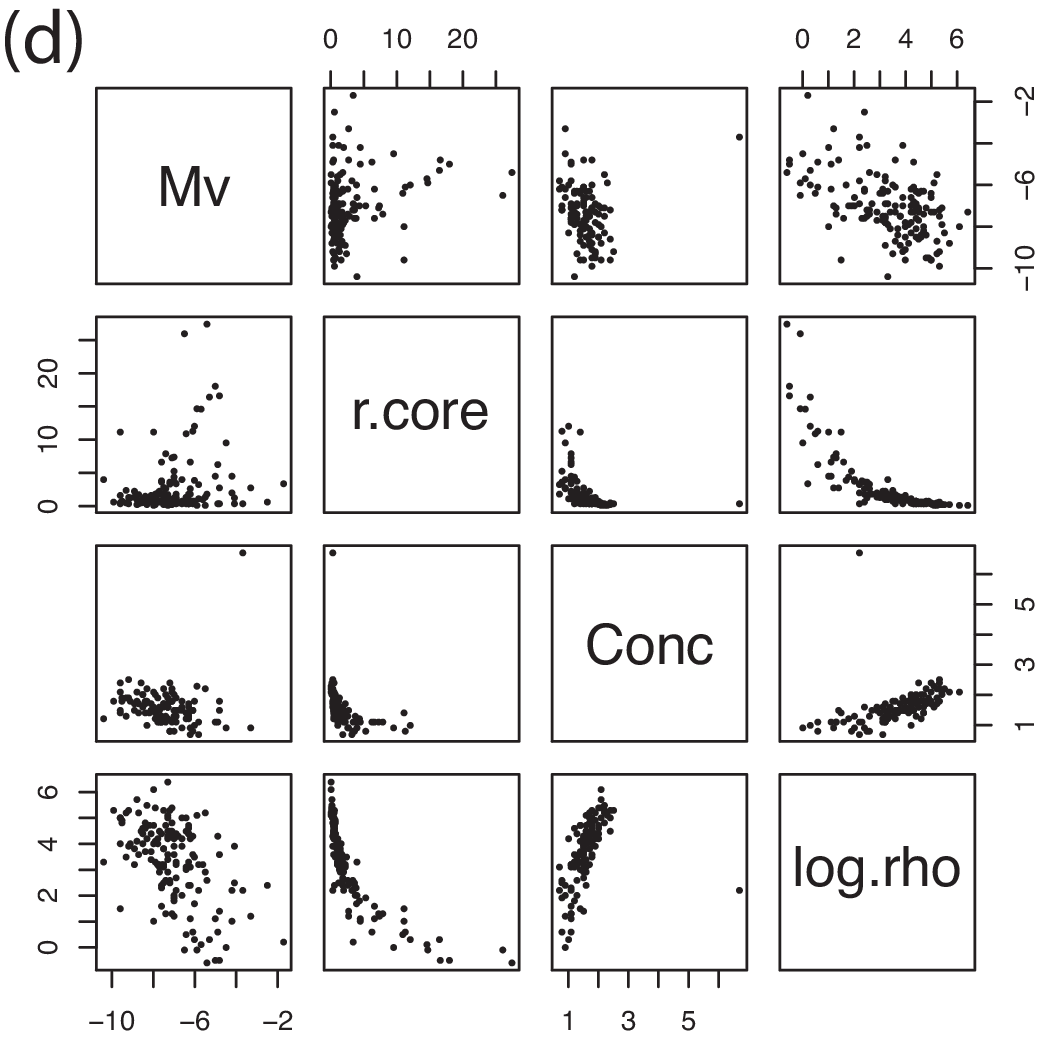}
\includegraphics[width=0.40\textwidth]{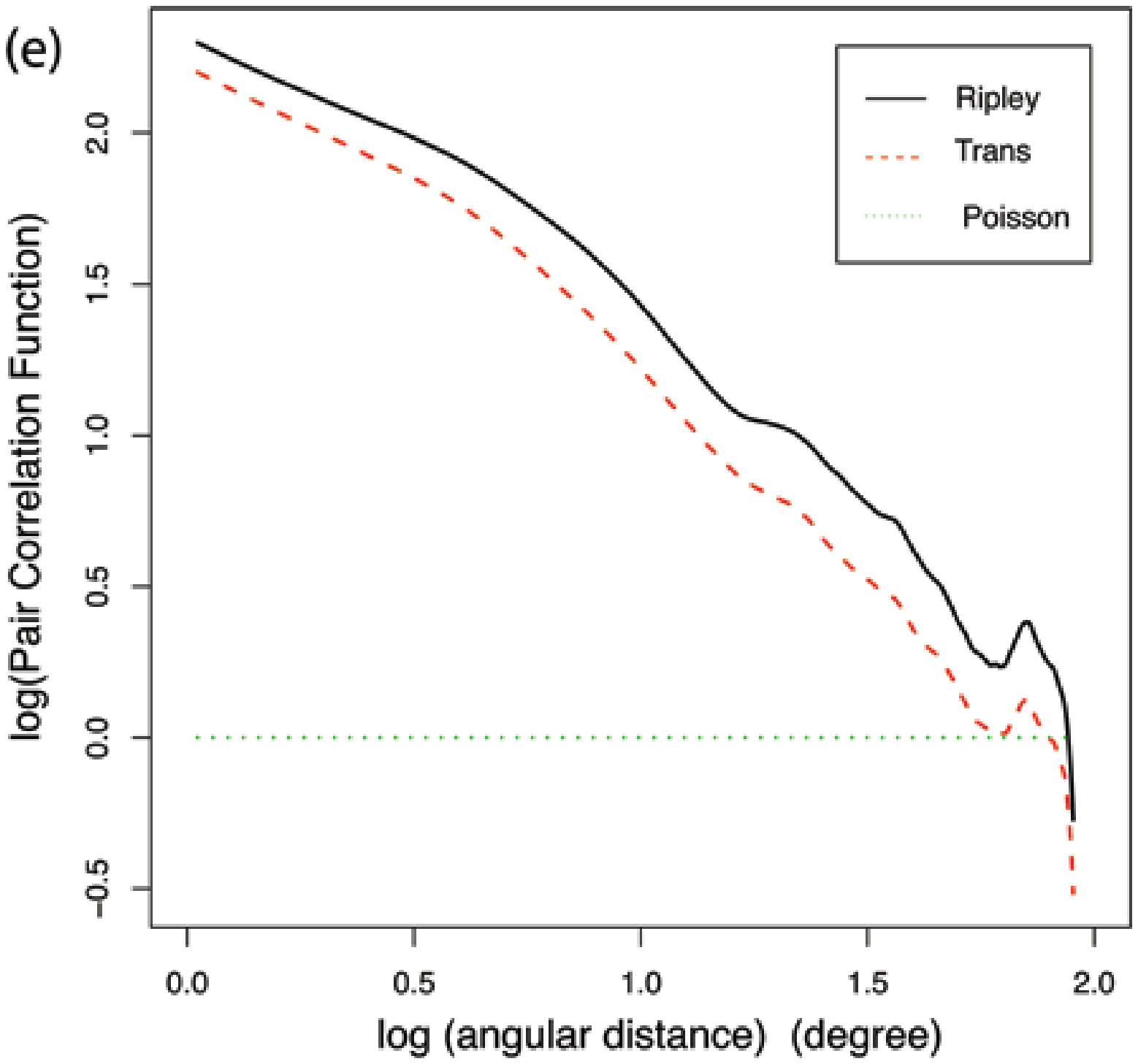} \\
\includegraphics[width=0.40\textwidth]{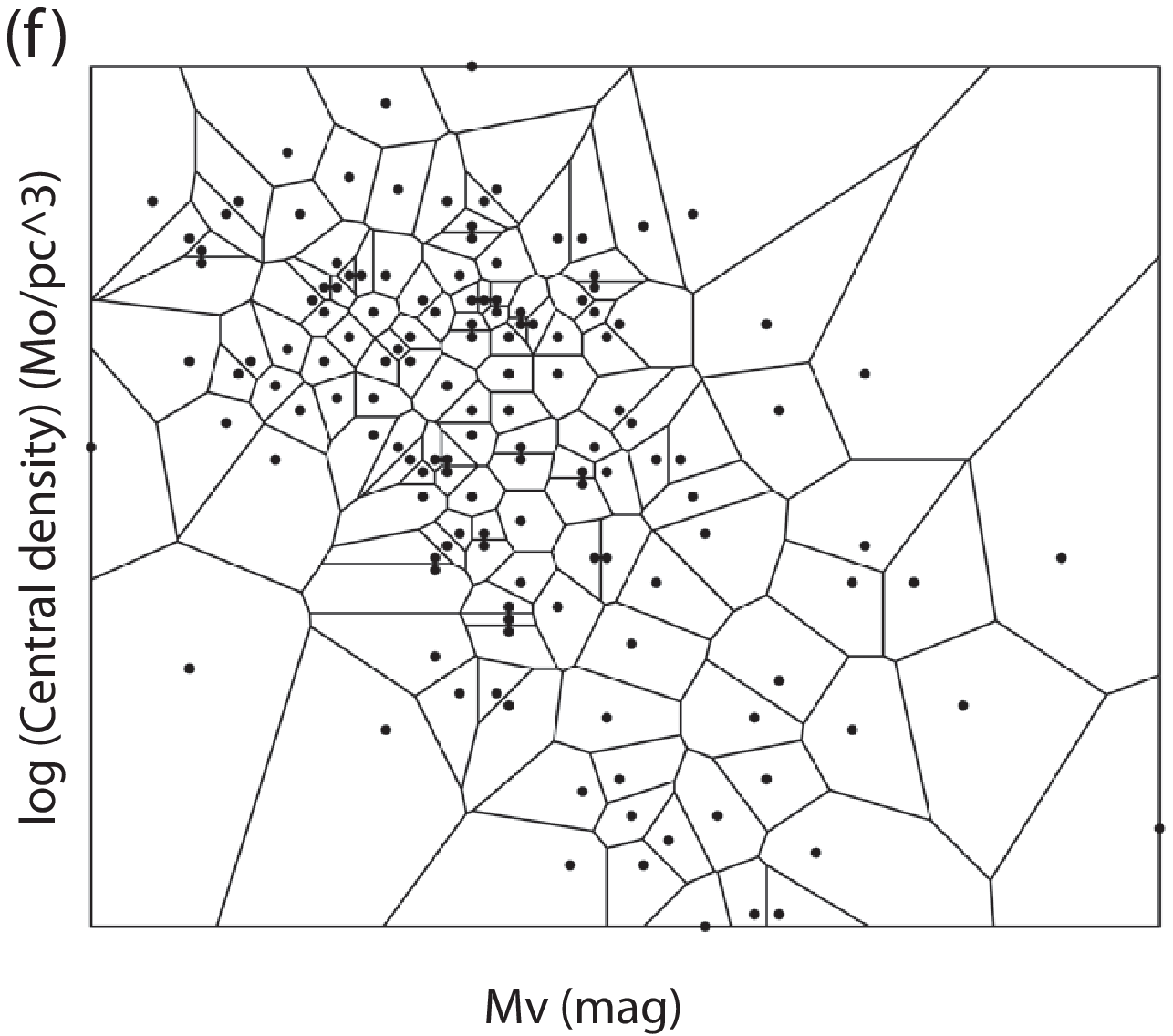}
\includegraphics[width=0.5\textwidth]{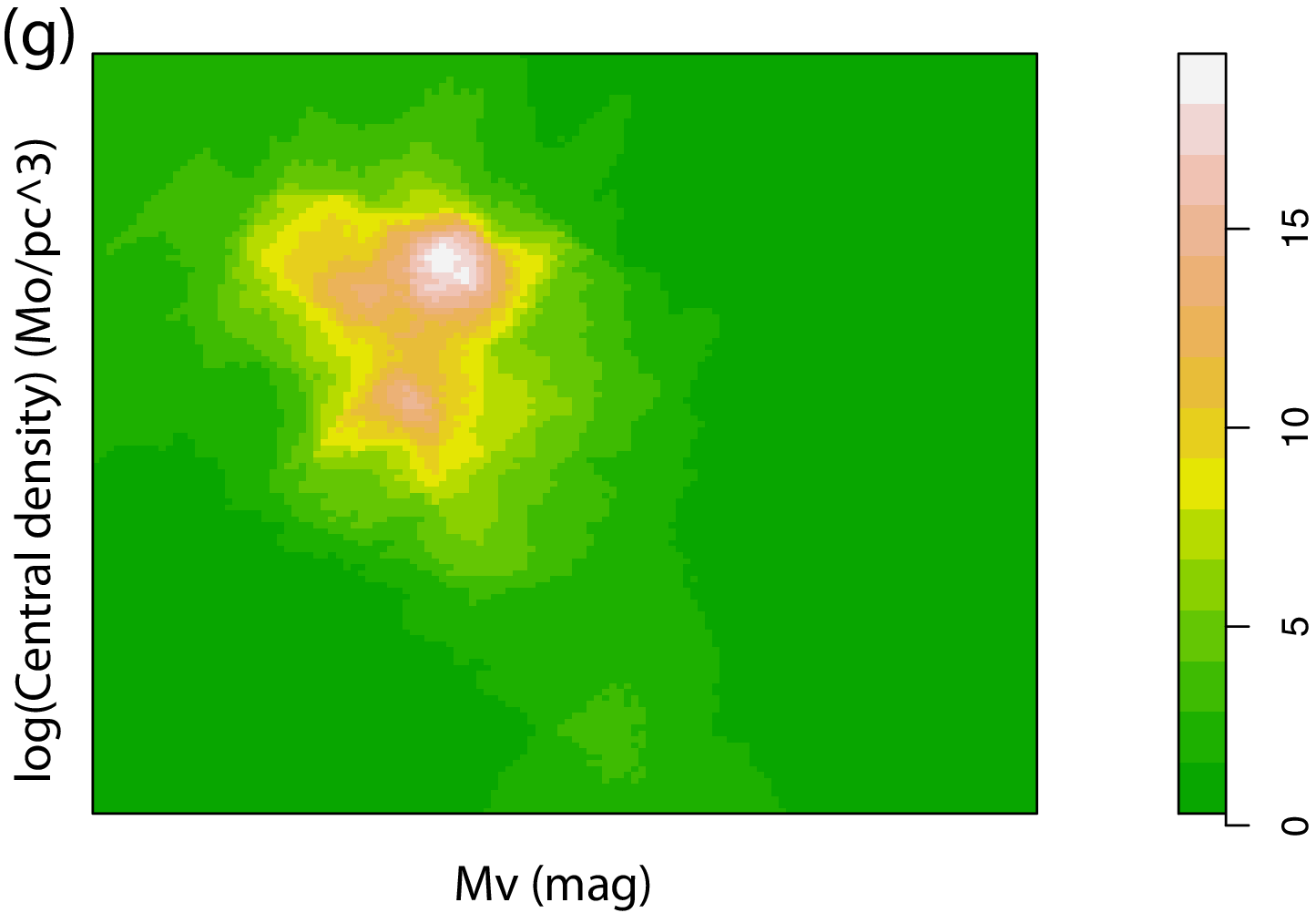}
\end{figure}

\subsection{Estimating distributions from data} \label{funct_dists}

{\bf Histogram} and {\bf Averaged shifted histograms} ~~ See \S \ref{funct_plots}.

{\bf Empirical distribution function} ~~ This provides a sequence of plots of the normalized cumulative step-function of a single variable with 95\% confidence bands based on the Kolmogorov-Smirnov statistic using the {\it sfsmisc} {\bf  CRAN} package \citep{Machler11}. [{\it ecdf.ksCI}] 

{\bf Kernel density estimation: 1-dimensional, fixed window width} ~~ This function convolves the data with a Gaussian function.  Here a constant bandwidth selected by the user is used; if no bandwidth is specified, a default optimized to a Gaussian distribution for the data, $h = (\frac{4}{3n})^{1/5} \sigma$ where $n$ is the number of data points and $\sigma$ is the sample standard deviation.  A rug plot of the input data and 95\% confidence bands around the smooth estimator are shown. This is produced with {\bf  CRAN} package {\it sm} \citep{Bowman97, Bowman10}.  [{\it sm.density, plot, lines, rug}] {\sc SAMP} 

{\bf Kernel density estimation: 1-dimensional, adaptive window width} ~~ This is a sophisticated robust $L$-estimator that applies a Gaussian kernel with bandwidth adapted to the local concentration of data points implemented in the {\bf  CRAN} package {\it quantreg} \citep{Portnoy89, Koenker12}.  The user can adjust a parameter that determines the sensitivity of the local bandwidth to local variations.  [{\it akj, seq, plot, rug}] {\sc SAMP} 

{\bf Kernel density estimation: 2-dimensional, fixed window width} ~~ This function convolves the data with a Gaussian function of fixed bandwidth and produces a colorful 3-dimensional surface plot that can be interactively rotated \citep{Bowman97}.  Confidence bands are available but not plotted. [{\it sm.density}] {\sc SAMP} 

{\bf Fitting distributions to data} ~~  This performs a maximum
likelihood fit of a statistical distribution to a univariate
dataset.  Fifteen distribution are provided: beta, cauchy,
chi-squared, exponential, F, gamma, geometric, log-normal,
logistic, negative binomial, normal, Pareto (power law), Poisson,
t and Weibull.   The best-fit parameters are obtained numerically
except for the normal, log-normal, geometric, exponential and
Poisson distributions where analytic solutions are used.  Output
includes best fit parameters, their estimated standard errors
from the Fisher information matrix, and the log-likelihood.  The
{\bf R} implementation is described by \citep{Venables02}. The
only exception is Pareto distribution, for which VOStat uses its
own code. [{\it fitdistr}]  

\subsection{Comparing data with distributions} \label{funct_compare}

{\bf Normal plot} ~~ This function provides a quantile-quantile (Q-Q) plot of a univariate dataset compared with the normal (Gaussian) distribution.  If the distribution agrees well with a normal, then a variety of classical statistics can be applied. [{\it qqnorm, qqline}] 

{\bf Tests for normality} ~~ Several hypothesis tests are provided to test whether a univariate dataset is consistent with a normal (Gaussian) distribution.  These include the Shapiro-Wilk test in base {\bf R}, and the Lilliefors (specialized Kolmogorov-Smornov), Cramer-von Mises, Anderson-Darling, Pearson (chi-square),  and Grubbs tests from {\bf CRAN} package {\it nortest}. [{\it shapiro.test, lillie.test, cvm.test, ad.test, pearson.test, sf.test}]   

{\bf One-sample KS test} ~~  A selected univariate sample is compared to a uniform, exponential or normal distribution (with user-specified parameters).  The two cumulative distribution functions are shown with the differences, and the Kolmogorov-Smirnov 1-sample test is applied.  [{\it ks.test, seq, punif, pexp, pnorm, ecdf}]
 
 {\bf Anderson-Darling test} ~~  As in the previous function, a univariate sample is tested against a uniform, exponential or normali distribution using the Anderson-Darling test.  This is more sensitive than the Kolmogorov-Smirnov test for multiple small-scale deviations and differences at the edges of the distribution. The function, implemented in {\bf CRAN} package {\it adk}, permits $k$-sample comparisons, but this is not implemented in VOStat \citep{Scholz08}.  [{\it adk.test}]
 
 {\bf Two-sample KS test} ~~ The distributions of two variables in a multivariate dataset are compared using the two-sample Kolmogorov-Smirnov test.  The user selects whether a one-sided or two-sided test is desired. [{\it ks.test}]
 
 {\bf Chi-square two-sample test} ~~  Two univariate samples of grouped (binned) count data are combined into a contingency table, and their consistency is estimated using Pearson's chi-squared test.  The two samples must have the same number of bins.  [{\it table, cbind, chisq.test}]

\subsection{Some standard tests} \label{funct_tests}

{\bf One-sample $Z$ test} ~~  This is a hypothesis test that compares the mean of a chosen univariate dataset with a user-specified value.  It applies when the dataset is approximately normal (Gaussian) distribution, the variance is known, and the sample is sufficiently large.  [{\it nrow, pnorm}]

{\bf One-sample $t$ test} ~~ This is a similar hypothesis test for the mean but does not require prior specification of the variance.  [{\it t.test}]

{\bf Paired $t$ test} ~~ This is a similar hypothesis that the means of two univariate samples have the same mean (or with a specified offset).  Note the assumption of Gaussianity.  [{\it t.test}]

\subsection{Regression} \label{funct_regress}

{\bf Linear regression} ~~  Multivariate linear regression is performed using {\bf R}'s important {\it lm} (linear modeling) function.  Its use is described in detail by \citet{Sheather09}.  The user specifies a single response variable (optionally weighted by  measurement errors) and one or more independent variables (without measurement errors). The fitted regression coefficients and their confidence intervals are produced, along with diagnostic residual plots to assist in evaluating the quality of the fit.   {\it (lm, length, for, plot)}  {\sc SAMP} 

{\bf Nadaraya-Watson local regression} ~~  This is a nonparametric bivariate kernel regression technique to estimate conditional expectation of the random variable $Y$ on $X$.  It can find nonlinear relationships of unknown functional form.  A constant bandwidth for a Gaussian kernel derived by Nadaraya and Watson in 1964 is used.  The resulting plot shows the bivariate scatter plot, local regression curve, and 95\% confidence intervals obtained from bootstrap resamples.    An example is shown in Figure~\ref{multivar_plots.fig}c.  The algorithm is implemented in {\bf CRAN} package {\it np} \citep{Hayfield08}. [{\it npregbw, npplot, points}] 

{\bf Major axis/orthogonal regression} ~~ While ordinary linear regression requires specification of a `response' or dependent variable, astronomers sometimes seek intrinsic relationships that treat random variables in a symmetrical fashion. VOStat implements three symmetrical bivariate least-squares linear regression fits using the {\bf CRAN} package {\it lmodel2} \citep{Legendre08}.   Astronomical applications of symmetric regression lines are described in \citet{Feigelson92}. 

{\bf LOESS regression}  ~~  A popular method for local polynomial regression fitting based on weighted least squares fitting of linear functions locally around each data point \citep{Cleveland94}.  A local parabola is used here, and the user specifies a smoothing bandwidth parameter.  [{\it loess, {\sc SAMP}, plot, order, lines}] {\sc SAMP}

{\bf Linear quantile regression} ~~   Quantile regression is a sophisticated method that fits, for sufficiently large datasets, linear and nonlinear relationships to the median or other quantile of the response variable conditioned on the independent variable \citep{Koenker05}.  This is valuable if the scatter about the relationship is asymmetrical or otherwise non-Gaussian.  The calculation is made by {\bf CRAN} package {\it quantreg} \citep{Koenker12}. 

{\bf Robust regression using an $M$-estimator} ~~ A method of iterative least squares multivariate regression where the influence of outliers are down-weighted using Huber's $\psi$ function \citep{Huber81}.  The VOStat implementation is restricted to linear fits, but a much wider range of functions is permitted in {\bf R}.  [{\it rlm, plot}] {\sc SAMP}

{\bf Robust regression using trimmed least squares} ~~  Another linear robust regression method that removes, rather than downweights, outlying points \citep{Rousseeuw03}.  Here the the sum of the quantile smallest squared residuals is minimized.  [{\it ltsreg, plot}] {\sc SAMP}

\subsection{Multivariate methods} \label{funct_multivar}

{\bf Pairs plot} ~~  See \S\ref{funct_plots}.

{\bf Linear regression} ~~ See \S\ref{funct_regress}. 

{\bf Principal component analysis} ~~ An important procedure to find the most important linear combinations of variables that account for the variance in the dataset.  [{\it princomp, na.omit, plot}] {\sc SAMP}

{\bf Canonical correlation analysis} ~~ A multivariate procedure that finds two subsets of variables whose linear combinations have maximum correlation with each other.  [{\it cancor, no.omit}] 

{\bf Model based clustering} ~~ This is an important method to find structures in multivariate data assuming they follow multivariate normal (Gaussian) distributions. These can be shaped like spheres, ellipsoids, pancakes or cigars in $p$-dimensions. This is called a `normal mixture model' \citep{McLachlan00}.  Parameters are obtained by maximum likelihood estimation using the EM Algorithm with model selection (i.e., choice of number of clusters) based on the Bayesian Information Criterion.   This well-known code is in {\bf CRAN} package {\it mclust} \citep{Fraley06}.  The VOStat code produces a plot of BIC $vs.$ number of clusters, a pairs plot with symbols reflecting cluster membership, and tabular output giving the multivariate means and variances of each cluster.  [{\it mclutBIC, plot, summary}] {\sc SAMP}

{\bf Hierarchical clustering} ~~ Nonparametric hierarchical clustering is based on agglomerating proximate data points into increasingly larger groupings using a chosen metric and cluster criterion.  VOStat assumes a Euclidean metric, so standardization of variables may be desirable.  Clustering criteria include: Ward's, single linkage, average linkage, complete linkage, McQuitty's criterion, median and centroid.  Note that `single linkage' clustering is commonly called the `friends-of-friends' algorithm in astronomy, but is not recommended due to its propensity to `chain' together distinct structures \citep{Everitt11}.    [{\it dist, hclust, plclust, cophenetic, cor, cutree}] {\sc SAMP}

\subsection{Spatial methods}\label{funct_spatial}

{\bf Variogram} ~~ A variogram is a function measuring the spatial dependence of a random field that is sampled at data points in two dimensions.  Three variables are involved: the variable whose spatial pattern is of interest, and two variables representing location.  The plot shown here gives the variance of the difference between the intensity of the process at two locations separated by a distance in the spatial plane. The calculation assumes the point process is stationary (same behavior at all locations) and isotropic.  The plots of temperature fluctuations as a function of angular scale for the cosmic microwave background, and of velocity dispersion as a function of physical scale in molecular clouds are astronomical examples of variograms.  It is implemented here in the {\bf CRAN} package {\it gstat} \citep{Pebesma04}.  [{\it variogram, plot}]

{\bf $k$-nearest neighbors and Moran's $I$ test} ~~ The
$k$-nearest neighbors method useful for forming an idea about the
topology of a point process. Each point is connected to its $k$
nearest neighbors. Moran's test can check for spatial
autocorrelation of a variable with respect to this topology.  It
gives the probability that the autocorrelation is consistent
with complete spatial randomness.  [{\it  spdep, RANN}]

{\bf Ripley's $K$ and related measures} ~~ Astronomers commonly examine the 2-point correlation function (known in statistics as the `pair correlation function') for clustering of a stationary point process, showing the strength of spatial clustering as a function of distance.  Statisticians commonly use the integral of the 2-point correlation function to avoid arbitrary binning; this is known as Ripley's $K$ function. The plot provided by VOStat shows the $K$ function with two corrections of edge effects, and a theoretical curve assuming complete spatial randomness.  Also shown are Besag's $L$ (the $K$ function rescaled to be horizontal in the null case of random distributions), Baddeley's $J$ (a different function that is less sensitive to edge effects), and the pair correlation function.  Computations are performed with the {\bf CRAN} package {\it spatstat} \citep{Baddeley05}.  A pair correlation function is shown in Figure~\ref{multivar_plots.fig}e.  [{\it as.matrix, owin, as.ppp, Kest, Lest, Jest, pcf, plot}]  

{\bf Voronoi tessellation and adaptive 2D kernel density estimation} ~~  The Voronoi (or Dirichlet) tessellation is a division of the space into distinct regions around each data point defined such that each tile contains the locations closer to that point than any other point.  The result is  the division of the space into polygons around each point. The inverse of the polygonal area is a measure of the local density of points, and is used to create a nonparametric adaptive smoother of the distribution of points. VOStat provides both a 2-dimensional grey-scale map and a 3-dimensional perspective plot of the smoothed distribution that can be interactively rotated.  These operations are computed with {\it spatstat}.  A Voronoi tessellation and its associated adaptively smoothed distribution are shown in Figure~\ref{multivar_plots.fig}f-g.   [{\it as.matrix, owin, dirichelet, adaptive.density, plot, surf3d}]

\subsection{Directional statistics} \label{funct_direct}

{\bf Plotting directional data} ~~ A circular plot is done to
provide an idea about the general shape of the data. It is an
anlog of barplot or histogram for directional data. A directional data plot is shown in Figure~\ref{multivar_plots.fig}a. 

{\bf Summarizing directional data} ~~ Since directional data deal
with angles, one has to be careful about $2\pi$ radians wrapping
back to $0.$ VOStat computes the mean direction as well as the
mean resultant length. [{\it circular}] 

{\bf Kernel density estimation} ~~ This tools computes kernel
density estimate from directional data. [{\it circular}] ~~ SAMP

{\bf Fitting a von Mises distribution} ~ The von Mises family of
distributions is like an analog of the Gaussian family for
directional data. Given a circular data set this tool performs
maximum likelihood fitting of a von Mises distribution to it.[{\it CircStats}]

{\bf Testing goodness-of-fit with a von Mises distribution} ~~
Usually fitting a distribution to a data is not enough until we
know if the fit is adequate. Watson's test checks
goodness of fit of a given data set with a fitted von Mises
distribution. It requires bootstrapping. [{\it CircStats}] 

{\bf Testing goodness-of-fit with a uniform distribution} ~~ Same
as above but with uniform distribution in place of von Mises
distribution. Four different tests are performed: Raleigh test,
Watson test, Kuiper test and Rao's spacing test. [{\it CircStats}]

\begin{figure}[h]
\centering
\includegraphics[width=0.7\textwidth]{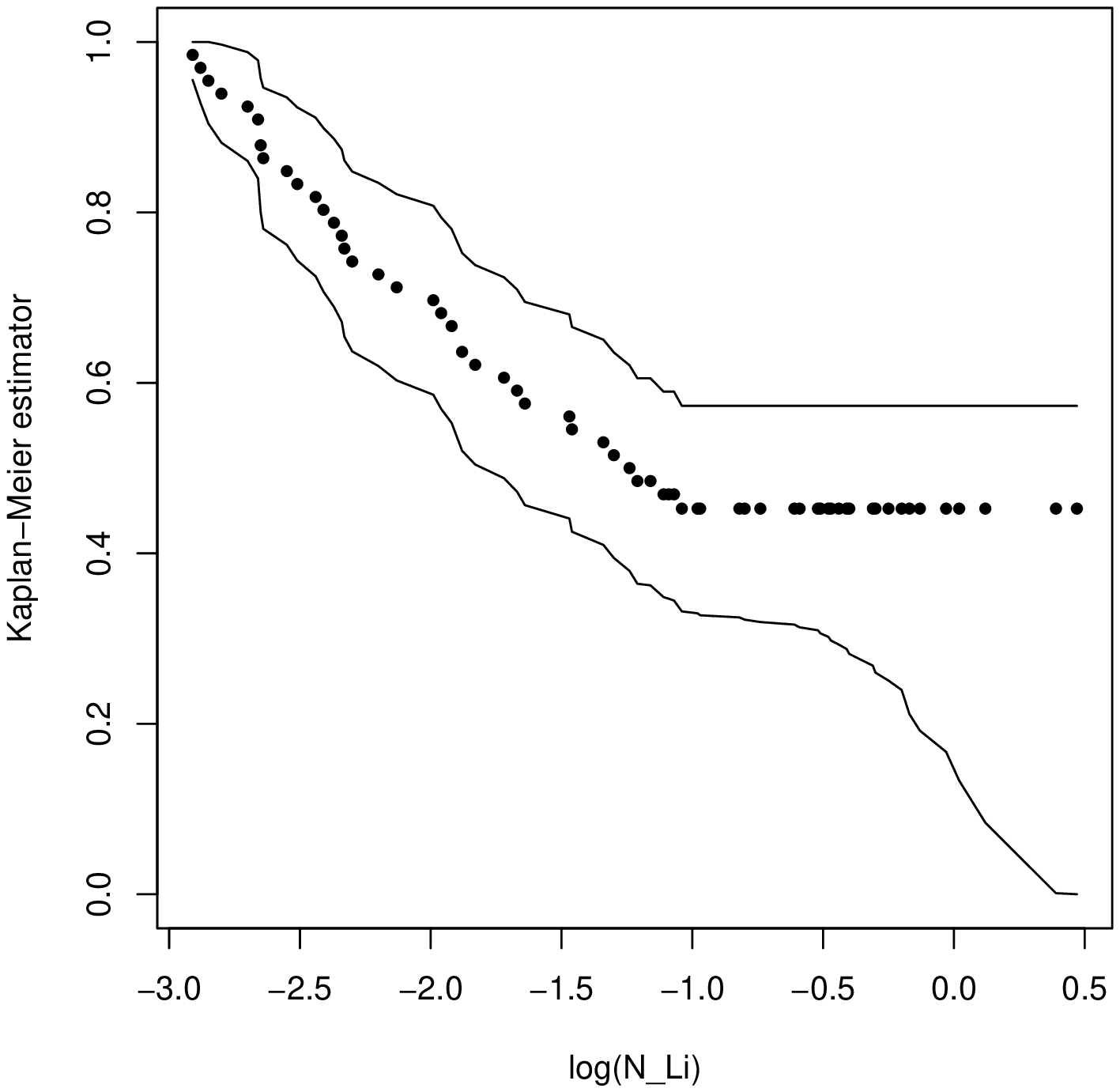}
\caption{VOStat plot for censored data (upper limits).  Kaplan-Meier estimator (with 95\% confidence bands) for lithium abundances of stars studied by \citet{Santos02}. Thirty-eight values are detected and thirty are upper limits. Data and analysis are described in \citet{Feigelson12}. }
\label{censor_plots.fig}
\end{figure}

\subsection{Survival analysis} \label{funct_surv}

{\bf Maximum likelihood Kaplan-Meier estimate of the survival curve for censored data} ~~ Astronomers often observe objects without detection, giving rise to upper limits of the observed property.  In statistics, these limits are called left-censored data points, and are treated with methods from `survival analysis'.  The Kaplan-Meier estimator is the unique maximum likelihood estimator of a randomly censored univariate sample.  Confidence intervals are based on Greenwood's formula.  A VOStat plot of a Kaplan-Meier estimator is shown in Figure~\ref{censor_plots.fig}.  [{\it Surv, surfit, plot, lines}] 

{\bf Tests for censored data} ~~ Two-sample tests to compare different univariate samples with censoring.  Generalizations of the Kolmogorov-Smirnov, Cramer-von Mises, and Anderson-Darling tests are applied.  Here, censoring patterns need not be random or the same in the two samples.  [{\it survdiff}] 

{\bf Maximum likelihood Lynden-Bell-Woodroofe estimator} ~~  While censoring occurs when known objects are not detected in some property, truncation occurs when an unknown number of unknown objects are not detected due to sensitivity limitations.  Analogous to the better-known Kaplan-Meier distribution for censoring, the Lynden-Bell-Woodroofe estimator gives the unique maximum likelihood estimator for a randomly truncated univariate variable \citep{Lynden-Bell71, Woodroofe85}.  Useful for estimation of luminosity functions in flux-limited astronomical surveys,  its mathematical properties are often better than heuristic functions like the widely-used $1/V_{max}$ estimator with arbitrary binning \citep{Schmidt68}.

\begin{figure}
\centering
\includegraphics[width=0.45\textwidth]{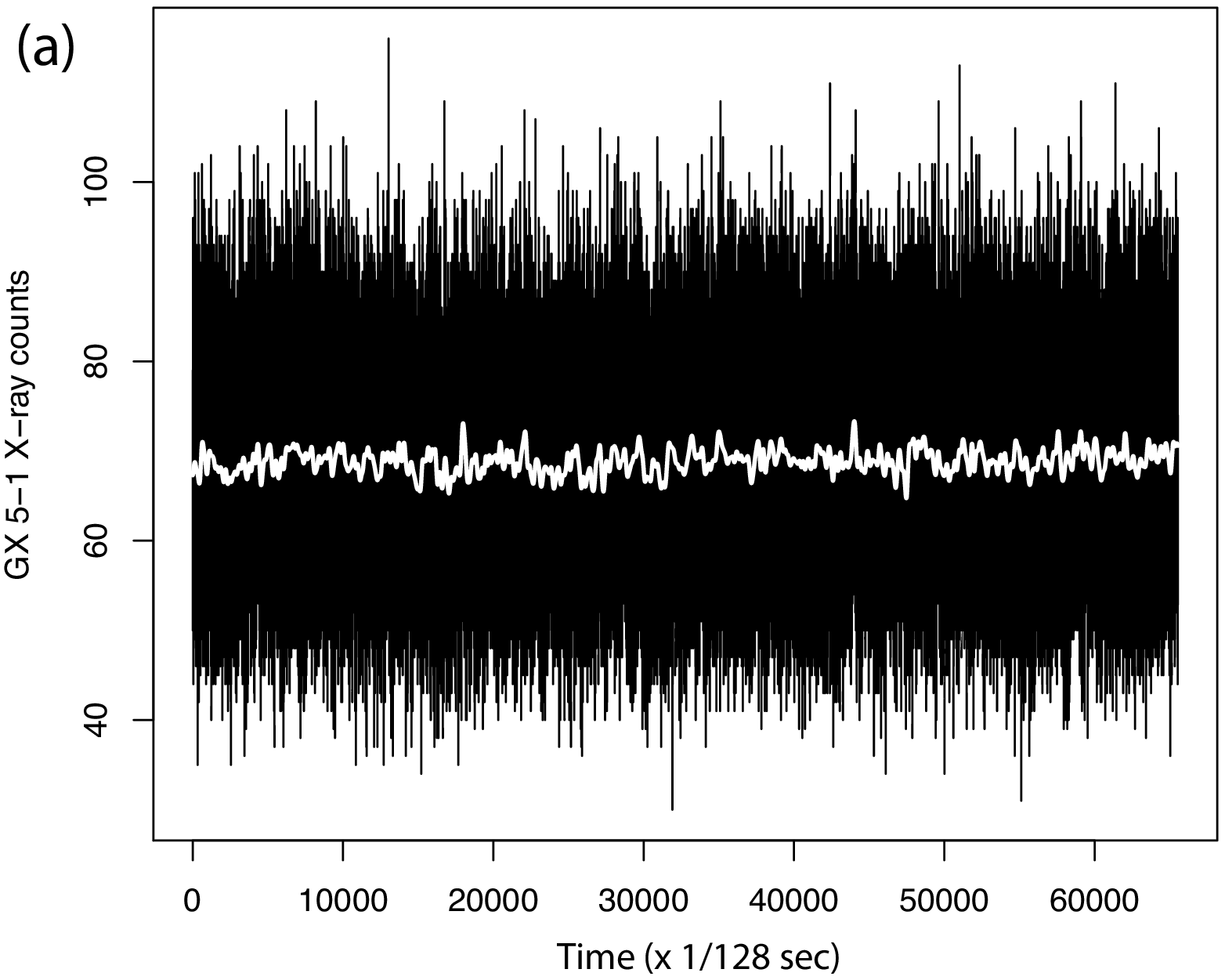}
\includegraphics[width=0.45\textwidth]{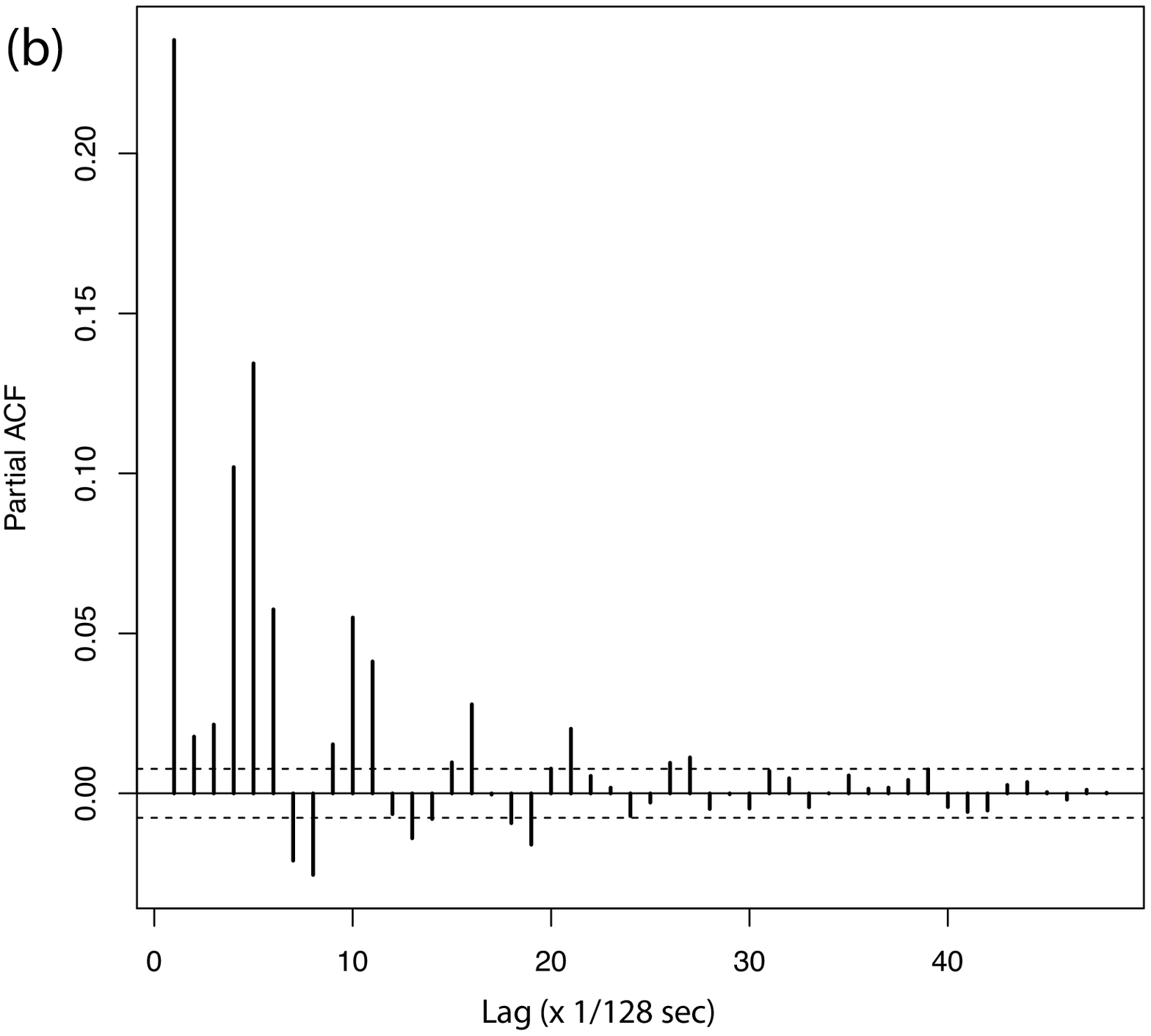} \\
\includegraphics[width=0.9\textwidth]{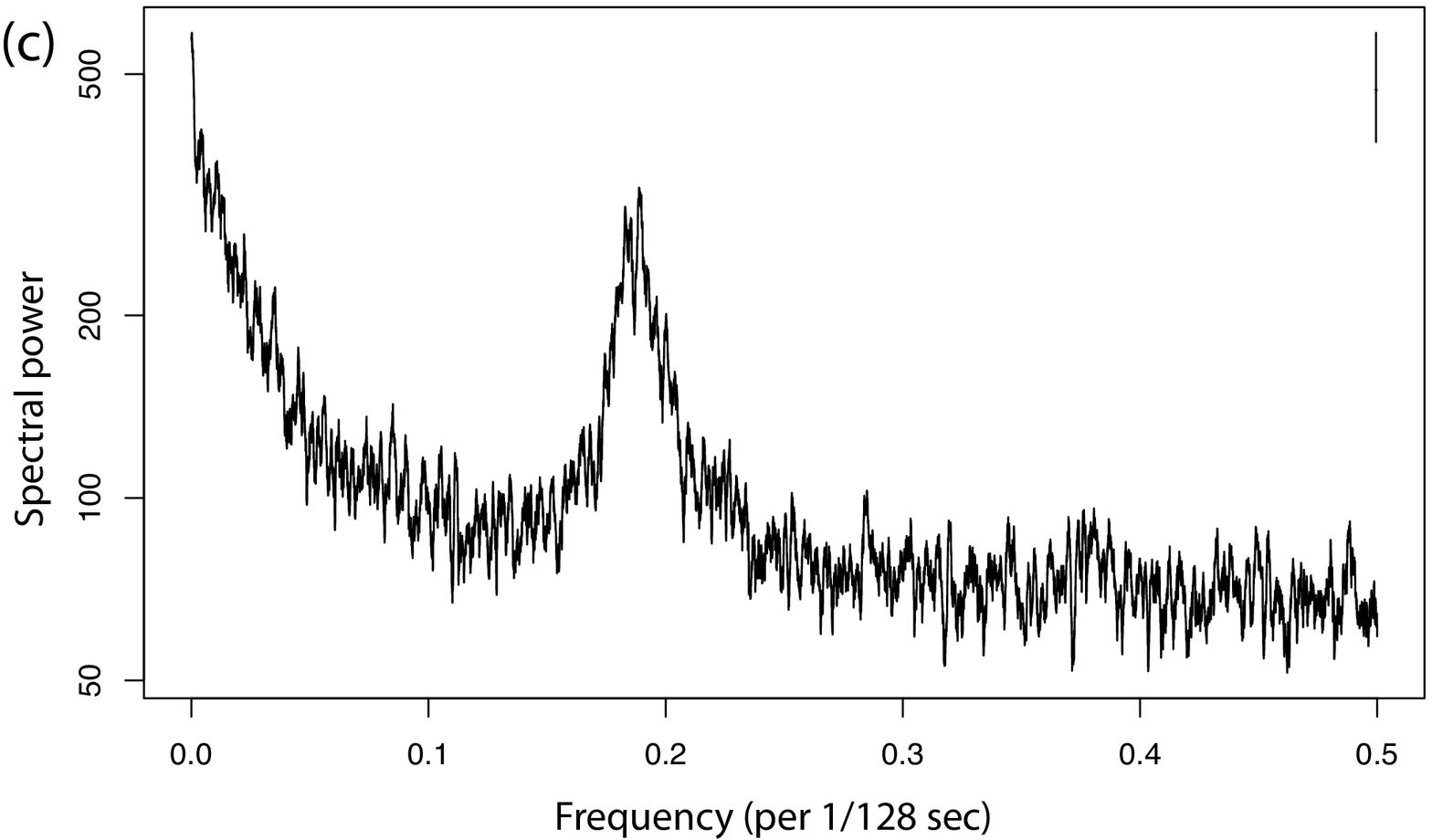}
\caption{VOStat plots for time series analysis of counts from the `quasi-periodic oscillator' X-ray source GX~5-1 \citep{Norris90}. (a) Time series plot with smoothing (bandwidth = 200 bins).  (b) Partial autocorrelation function showing complex structure, with 95\% confidence intervals assuming white noise. (c) Fourier periodogram with detrending, tapering and smoothing, with 95\% confidence interval assuming white noise.  Data and analysis are described in \citet{Feigelson12}. }
\label{TSA_plots.fig}
\end{figure}

\subsection{Time series analysis} \label{funct_TSA}

Nearly all time series analysis in {\bf R} and {\bf CRAN} requires regularly spaced data without gaps.  These methods are thus inappropriate for many irregularly spaced astronomical datasets. 

{\bf Time series plotting with smoothing} ~~ This VOStat function plots a univariate time series with Gaussian kernel smoothing based on user-provided bandwidth. A time series plot is shown in Figure~\ref{TSA_plots.fig}a. [{\it plot.ts, ksmooth, lines}]

{\bf (Partial) Autocorrelation Function (ACF and PACF)} ~~ Computation and plot of the autocorrelation function of the univariate time series; this is the correlation as a function of lag time.  The partial autocorrelation function removes the effects of intermediate lags. A PACF plot is shown in Figure~\ref{TSA_plots.fig}b.  [{\it acf, pacf}]

{\bf Autoregressive modeling} ~~ This is a maximum likelihood fit of the time series to a stochastic autoregressive model, where the order of the model is selected by the maximum of the Akaike Information Criterion (AIC), a penalized likelihood measure.  VOStat provides plots of AIC against order, and shows the periodogram of the resulting best-fit model.  [{\it ar, plot}]

{\bf Periodogram with smoothing option} ~~  This function provides a Schuster periodogram based on a Fourier transform of the time series, with Daniell (boxcar) smoothing. The time series is first detrended by a linear function, a 10\% taper is applied, and a 95\% confidence interval assuming white noise is shown. A periodogram of an astronomical time series is shown  in Figure~\ref{TSA_plots.fig}c.  [{\it spec.pgram}]

\subsection{Conclusion}

While many resources are applied to the reduction of data as it arrives from telescopes, VOStat is one of the relatively few resources available to assist the astronomer in the later stages of scientific analysis, such as the discovery of patterns in the observations and comparison of observed properties of celestial objects with astrophysical theory.  As eloquently phrased by \citet{Gregory05}:  \begin{quote}
The goal of science is to unlock natureÕs secrets. ... Our understanding comes through the development of theoretical models which are capable of explaining the existing observations as well as making testable predictions. ... Fortunately, a variety of sophisticated mathematical and computational approaches have been developed to help us through this interface, these go under the general heading of statistical inference.Õ  \end{quote}

VOStat provides a user-friendly, interactive, Web-based interface that applies several dozen statistical operations to a user-supplied dataset.  A unique difference from other statistical services is its integration with other interactive tools of the Virtual Observatory using the VO's SAMP communication system between applications.   Using VO tools, an astronomer can extract carefully chosen datasets from huge, widely distributed, multiwavelength databases, analyzing and enhancing the data in various ways within VOStat.  Some results from VOStat analysis can then be broadcast back to other VO applications through SAMP for visualization and further analysis.    

The engine behind VOStat is the remarkable {\bf R} statistical software system, the most comprehensive available public domain data analysis environment.  We emphasize that VOStat taps only a small fraction of {\bf R}, and a tiny ($<1$\%) fraction of the specialized {\bf  CRAN} packages.  Thus, we strongly encourage VOStat users to extend the on-line analysis with a more thorough interactive analysis on their home computer using {\bf R} and {\bf CRAN}.  Educational materials for learning {\bf R} are available at http://www.r-project.org and in many books, including one designed specially for astronomy \citep{Feigelson12}.  

\acknowledgments

This work is supported by the NSF `SI2-SSE' grant AST-1047586 (PI - G. J. Babu).


\end{document}